\documentclass[aps,prl,amsfonts,nofootinbib,supercriptaddress]{revtex4}

\usepackage{amssymb,amsmath}
\usepackage{graphicx}
\usepackage{epsfig}
\usepackage{citesort}

\newcommand{\beq}{\begin{equation}}
\newcommand{\eeq}{\end{equation}}
\newcommand{\bea}{\begin{eqnarray}}
\newcommand{\eea}{\end{eqnarray}}
\newcommand{\vc}[1]{{\textbf{#1}}}
\newcommand{\mc}[1]{\mathcal{#1}}
\newcommand{\etas}{\eta_s}

\begin{document}

\title{Supernovae data and perturbative deviation from homogeneity}

\author{Kari Enqvist\footnote{E-mail: kari.enqvist@helsinki.fi}}
\affiliation{Helsinki Institute of Physics, P.O.Box 64, FIN-00014
University of Helsinki, Finland; \\ Physics Department, P.O. Box 64,
FIN-00014 University of Helsinki, Finland}

\author{Maria Mattsson\footnote{E-mail:
maria.ronkainen@helsinki.fi}}

\author{Gerasimos Rigopoulos\footnote{E-mail:
gerasimos.rigopoulos@helsinki.fi}} \affiliation{Helsinki Institute
of Physics, P.O.Box 64, FIN-00014 University of Helsinki, Finland}

\begin{abstract}

\noindent We show that a spherically symmetric perturbation of a
dust dominated $\Omega=1$ FRW universe in the Newtonian gauge can
lead to an apparent acceleration of standard candles and provide a
fit to the magnitude-redshift relation inferred from the supernovae
data, while the perturbation in the gravitational potential remains
small at all scales. We also demonstrate that the supernovae data
does not necessarily imply the presence of some additional
non-perturbative contribution by showing that any
Lemaitre-Tolman-Bondi model fitting the supernovae data (with
appropriate initial conditions) will be equivalent to a perturbed
FRW spacetime along the past light cone.

\end{abstract}

\maketitle

\section{Introduction}

A basic principle which has enabled substantial progress in
cosmology, referred to as the Copernican Principle, is the
assumption that no observer has a privileged position in space.
Coupled with the observed isotropy of the universe around us, this
principle leads to the conclusion that the universe is homogeneous.
Although the consistency of the current cosmological models strongly
supports this idea, the Copernican Principle has never been
explicitly tested; even the possibility of such testing has only
been raised rather recently
\cite{Caldwell:2007yu,Clarkson:2007pz,Uzan:2008qp}. Even if the
Copernican Principle is in general valid, there is still the
possibility that it is mildly violated at some cosmologically
significant level and that observers could occupy inequivalent
spatial locations. In that case, their observations might mislead
them into inferences about the universe as a whole and its global
parameters which are incorrect.

Doubts about the validity of the Copernican Principle lie very much
at the heart of some of the explanations put forward to account for
the apparent acceleration implied by the Type I supernovae data
\cite{supernovae}. As is well known,  all cosmological data is
consistent with a homogeneous $\Lambda$CDM concordance model, which
assumes the existence of a cosmological constant $\Lambda$ that
makes up about 70\% of the current energy budget of the universe.
Equally well acknowledged is the fact that, from a theoretical point
of view, the incredibly small value of the required cosmological
constant, about $10^{-120}$ $M_{\rm pl}$, is a mystery. Moreover,
there is also a peculiar coincidence that $\Lambda$ appears to have
become important at the time when structures started to form. This
may be taken as a hint towards the possibility that inhomogeneities
could in some fashion be responsible for the apparent acceleration.
One school of thought attempts to explain it as the influence,
sometimes called backreaction, of averaging inhomogeneities on the
average expansion of the universe
\cite{Buchert:1999er,Buchert:2007ik,Rasanen:2006kp,Rasanen:2008it,Paranjape:2008ai,Paranjape:2008jc,Kolb:2008bn,Mattsson:2007qp,Mattsson:2007tj}.
Backreaction would be an inherently nonlinear effect due to the
nonlinearity of general relativity. In this case the Copernican
Principle could still hold in some statistical sense. Another
possibility is to abandon the Copernican Principle altogether and
assume that our location in the universe is exceptional, thereby
fooling us into thinking that a cosmological constant exists, while
in reality it is only a trick of the light. Such considerations have
mostly been put forward within the context of spherically symmetric
Lemaitre-Tolman-Bondi (LTB) models \cite{LTB1,Krasinski,Krasinski2,
Enqvist:2007vb}. Many authors have considered various possibilities
offered by LTB models \cite{LTBpapers}, which include
inhomogeneities in the matter distribution as well as
inhomogeneities in the expansion rate; an example is a dust universe
with a local underdensity \cite{LTBvoid} that seems to be able to
reproduce the SNIa observations.

There is an ongoing debate about the role of inhomogeneities,
particularly whether the effects of inhomogeneities are too large to
be describable within the standard perturbed FRW description. In
contrast to various model calculations, in the present paper we do
not make an attempt at model building and full-blown data fitting.
Rather, we address the generic question of whether the SNIa data
requires a deviation from the Copernican Principle that is
non-perturbative. To explore this issue, we consider an FRW
spacetime that has small perturbations in the gravitational
potential $\psi$ (although the density perturbations could well be
large). We assume an $\Omega = 1$ dust universe and, for simplicity,
spherically symmetric perturbations, and solve for $\psi$ and the
velocity field ${\bf v}$ such that the distance-luminosity relation
of the SNIa data is satisfied; this we do in Newtonian gauge. We
find that the SNIa data can be fitted with $\psi$ and ${\bf v}$
remaining small during the whole evolution, meaning that the
perturbed FRW framework works well and the Copernican Principle is
thus only weakly violated in the sense that there are only small
deviations from homogeneity. Furthermore, one might then argue that
any effects of averaging over these small inhomogeneous
perturbations, i.e. the backreaction, remain negligible
\cite{Ishibashi:2005sj}; due to the nonlinear nature of
backreaction, the effects of averaging can only arise at second
order in the perturbations and would hence be extremely small.
Indeed, if the spacetime metric of the real universe were a
perturbed FRW, it would imply that the backreaction is insignificant
and thus could not solve the dark energy problem
\cite{Paranjape:2008jc,Paranjape:2009zu}. However, we should note
that the backreaction issue is somewhat more involved, and refer the
reader to \cite{Paranjape:2009zu,Paranjape:2008ai,Paranjape:2008jc}
and the references therein. Indeed, in a nonlinear theory there
could arise additional non-perturbative contributions to metric
perturbations which are not manifest at any order of perturbation
theory performed about the FRW background but could be revealed by a
careful comparison of data and results obtained from perturbation
theory. As we will demonstrate, at least the supernovae data does
not necessarily imply the presence of some additional
non-perturbative contribution.

Finally, we contrast our result with LTB models, which under certain
conditions are known to be physically equivalent to perturbed FRW
models \cite{Paranjape:2008ai,Paranjape:2008jc,VanAcoleyen:2008cy},
although it has also been claimed that the LTB models capable of
accounting for the SNIa data cannot be obtained from a perturbed FRW
model \cite{Kolb:2008bn}. Here we show explicitly that such
conditions are fulfilled for realistic LTB models fitting the SNIa
data and thus conclude that an explanation of the SNIa data in terms
of (spherically symmetric) inhomogeneities does not require
non-perturbative effects that would not be already present in a
perturbed FRW description.

\section{A spherical perturbation}

Let us assume that the spacetime can be described by a perturbed FRW
model in the Newtonian gauge (vanishing anisotropic stress)
\beq\label{pertmetric}
ds^2=a^2\left[-(1+2\psi)d\eta^2+(1-2\psi)d\vc{x}^2\right]~. \eeq The
effect of linear perturbations on the luminosity distance has been
calculated in the past
\cite{Sasaki:1987ad,Barausse:2005nf,Bonvin:2005ps}. We follow here
the notation of Bonvin, Durrer and Gasparini \cite{Bonvin:2005ps},
who find the luminosity distance $d_L(\etas, \hat{\vc{n}} )$ of a
source at conformal time $\etas$ in a universe described by the
metric (\ref{pertmetric}) as given by \cite{Bonvin:2005ps}
\bea\label{perts1}
\frac{d_L(\etas, \hat{\vc{n}} )}{1+z}-\Delta\eta  &=&  \left(\vc{v}_o \cdot \hat{\vc{n}} - \psi_o\right)\Delta\eta -2\Delta\eta \vc{v} \cdot \hat{\vc{n}} + 2 \int\limits_{\etas}^{\eta_o}d\lambda \, \psi + 2 \Delta\eta \int\limits_{\etas}^{\eta_o}d\lambda \,  \hat{\vc{n}} \cdot \nabla \psi  \nonumber \\
&& + 2 \int\limits_{\etas}^{\eta_o}d\lambda
\int\limits^{\lambda}_{\eta_s} d\bar{\lambda}\,  \hat{\vc{n}} \cdot
\nabla \psi - \int\limits_{\etas}^{\eta_o}d\lambda
\int\limits^{\lambda}_{\etas} d\bar{\lambda}
\left(\bar{\lambda}-\etas\right) \left(\nabla^2\psi -
\hat{\vc{n}}\cdot \nabla \left(\hat{\vc{n}}\cdot\nabla \psi\right)
\right)~, \eea where $\vc{n}=-\hat{\vc{e}}_r$ refers to the spatial
direction of light propagating towards the observer, $\etas$ is the
conformal time at emission, and $\Delta\eta \equiv \eta_o-\etas$.
The integrands of the above integrals are taken to depend on
$\lambda$ ($\bar{\lambda}$) as well as on $\vc{x}=\vc{x}_o -
\hat{\vc{n}}\left(\eta_o-\lambda\right)$, i.e. they are evaluated on
the past light cone. The subscript ``$o$'' refers to the observer
today.

In Eq.\ (\ref{perts1}) $\etas$ refers to the background conformal
time corresponding to the emission of  light from the source.
However, in the presence of perturbations a fixed value of $\eta$
does not correspond to a spatially homogeneous value of $z$.
Furthermore, it is $z$, not $\eta$, which is measurable and by
considering any quantity as a function of redshift in effect one
slices the universe in slices of constant $z$, not $\eta$. From Eqs.
(56) and (57) of \cite{Bonvin:2005ps} we find for the luminosity
distance as a function of redshift the expression: \beq \frac{d_L(
z, \hat{\vc{n}} )}{1+z}-\Delta\eta = \frac{d_L( \etas, \hat{\vc{n}}
)}{1+z}-\Delta\eta + \left(\Delta\eta +
\frac{1}{\mc{H}_s}\right)\left[\psi_s-\psi_o +
\left(\vc{v}_o-\vc{v}_s\right)\cdot \hat{\vc{n}} + 2
\int\limits_{\etas}^{\eta_o}d\lambda \, \hat{\vc{n}}\cdot\nabla\psi
\right]\,, \eeq from which we obtain \bea\label{perts2} \frac{d_L(
z, \hat{\vc{n}} )}{1+z}-\Delta\eta &=&
\left(2\Delta\eta+\frac{1}{\mc{H}_s}\right)\left(\vc{v}_o \cdot
\hat{\vc{n}} - \psi_o \right) -
\left(3\Delta\eta+\frac{1}{\mc{H}_s}\right) \vc{v}_s \cdot
\hat{\vc{n}}
+\left(\Delta\eta+\frac{1}{\mc{H}_s}\right)\psi_s \nonumber \\
&&+ 2 \int\limits_{\etas}^{\eta_o}d\lambda \, \psi
+ 2 \left(2\Delta\eta+\frac{1}{\mc{H}_s}\right) \int\limits_{\etas}^{\eta_o}d\lambda \,  \hat{\vc{n}} \cdot \nabla \psi  \nonumber \\
&& + 2 \int\limits_{\etas}^{\eta_o}d\lambda
\int\limits^{\lambda}_{\eta_s} d\bar{\lambda}\,  \hat{\vc{n}} \cdot
\nabla \psi - \int\limits_{\etas}^{\eta_o}d\lambda
\int\limits^{\lambda}_{\etas} d\bar{\lambda}
\left(\bar{\lambda}-\etas\right) \left(\nabla^2\psi -
\hat{\vc{n}}\cdot \nabla \left(\hat{\vc{n}}\cdot\nabla \psi\right)
\right)~. \eea

Eq.\ (\ref{perts2}) can be used to determine the fluctuations of the
luminosity distance induced by a Gaussian perturbation
\cite{Bonvin:2005ps}, such as the one commonly assumed to arise from
inflation and describing structure formation in the concordance
cosmology. However, we will turn the reasoning the other way round
and ask: Assuming the lhs of Eq.\ (\ref{perts2}) is a measured
function from SNIa observations and that the underlying model for
the background universe is flat and matter dominated with $\Omega_M
= 1$ and $\Lambda = 0$, can we determine the perturbation $\psi$ (or
$\vc{v}$) needed to give the observed luminosity distance? In other
words, can a local gravitational perturbation fool us into thinking
that light has propagated in a universe with the estimated value for
the cosmological constant? We will answer this question by assuming
that the perturbation $\psi$ exhibits spherical symmetry for
simplicity. As discussed in the Introduction, we are interested in
the case where the background can still be described by the FRW
metric. Furthermore, assuming linear theory, one can relate the
velocity field $\vc{v}$ with respect to the Einstein-de Sitter
background and the Newtonian potential by \beq\label{velocity}
\vc{v}=-\frac{2}{3\mc{H}}\nabla \psi\,. \eeq Our goal then is to
determine $\vc{v}$ and $\psi$, given the lhs of (\ref{perts2}) from
observations (The $d_L - z$ relation corresponding to an
$\Omega_\Lambda\simeq 0.7$, $\Omega_M\simeq 0.3$ universe).

It is convenient to derive a second order equation for the radial
velocity $v_r$, by acting with two derivatives along the past light
cone $\frac{d}{d\eta} \equiv \partial_\eta + \hat{\vc{n}}\cdot
\nabla$, assuming that it is the dominant component. Indeed, if the
radial component is to be able to account for the observed $d_L-z$
relation, it will have to be be much larger than the other velocity
components related to the standard small Gaussian density
fluctuations. Given $v_r$ the potential can be determined via Eq.\
(\ref{velocity}).

Let us now proceed in the manner just described. Taking two
derivatives of (\ref{perts2}) we obtain \beq\label{basic1}
\frac{d^2}{d\etas^2}\left[\frac{d_L}{1+z}-\Delta\eta \right] =
\left(3\Delta\eta + \frac{1}{\mc{H}_s}\right) \frac{d^2}{d\etas^2}
v_{sr} -\frac{3\mc{H}_s}{2}\left(7 \Delta\eta +
\frac{13}{3\mc{H}_s}\right) \frac{d}{d\etas} v_{sr} +
\frac{15\mc{H}_s^2}{4}\left(\frac{7}{5}\Delta\eta +
\frac{3}{\mc{H}_s}\right) v_{sr}~, \eeq where $v_{sr} \equiv -
\vc{v}_s\cdot\hat{\vc{n}}$. To derive this result we have used the
following: For a matter dominated universe
$\frac{d}{d\eta}\frac{1}{\mc{H}} = \frac{1}{2}$ and $\partial_\eta
\psi = 0$. The latter relation allows us to write
$d\psi/d\eta=\hat{\vc{n}} \cdot \nabla \psi = -
\frac{3\mc{H}}{2}\hat{\vc{n}}\cdot \vc{v}$. To eliminate the
$\nabla^2\psi$ term we assumed that $v_r$ is dominant, i.e. that the
main component of the peculiar velocity is radial. Thus \beq
\nabla\cdot\vc{v} =
\left(\frac{2}{r}+\hat{\vc{e}}_r\cdot\nabla\right)v_r =
\left(\frac{2}{r}+\frac{\partial}{\partial \eta} -
\frac{d}{d\eta}\right)v_r\,. \eeq Furthermore, \beq
\frac{3\mc{H}}{2} v_r= -\frac{\partial}{\partial r}\psi\,. \eeq
Since $\partial_\eta\psi=0$, we have that \beq 0=\partial_\eta\left(
\frac{3\mc{H}}{2} v_r \right)\,, \eeq from which we get
($r=\Delta\eta$) \beq\label{nablasquared}
  -\Delta\eta\,\nabla^2 \psi_s = \frac{3\mc{H}^2}{2} \left(\frac{\Delta\eta}{2}+ \frac{2}{\mc{H}}\right)v_{sr} - \frac{3\mc{H}\Delta\eta}{2} \frac{d}{d\etas}v_{sr}\,.
\eeq

We can now express Eq.\ (\ref{basic1}) in terms of the redshift z
using \beq\label{etazeta} \mc{H}=H_0\sqrt{1+z}\,,\quad
\text{and}\quad d\eta = - \frac{dz}{H_0\left(1+z\right)^{3/2}} \,
\Rightarrow \, \Delta\eta
=\frac{2}{H_0}\left(1-\frac{1}{\sqrt{1+z}}\right)~. \eeq We obtain
\beq\label{basic2}
\left(6-\frac{5}{\sqrt{1+z}}\right)(1+z)^3\frac{d^2v_{sr}}{dz^2}  +
\frac{3}{2}\left(20 -
\frac{44}{3\sqrt{1+z}}\right)(1+z)^2\frac{dv_{sr}}{dz} +
\frac{3}{4}\left(14+\frac{1}{\sqrt{1+z}}\right)(1+z)v_{sr} =
\frac{1}{H_0}\frac{d^2}{d\etas^2}\left[\frac{d_L}{1+z}\right] \eeq
This is our main result and can be seen as the equation which
defines the perturbation in the (radial) peculiar velocity given an
observed $d_L - z$ relation, acting as a source on the rhs. Given
appropriate boundary conditions, its solution determines the
velocity needed to give rise to the observed $d_L - z$ curve. We
stress again that the rhs is zero for a matter dominated FRW
universe and hence any deviation from zero is attributed to a
perturbation. Thus, the physically relevant boundary conditions for
our case are $v_r(z=0)=\frac{d}{dz}v_r(z=0) = 0$. Allowing for
curvature in the background model would bring complications to the
perturbation theory and thus would require a separate analysis, so
for simplicity and also motivated by the observations we restrict
ourselves to the flat model.

Using Eq.\ (\ref{etazeta}) we can calculate the rhs of Eq.\
(\ref{basic2}) to find \beq
\frac{1}{H_0}\frac{d^2}{d\etas^2}\left[\frac{d_L}{1+z}\right] = H_0
\left[(1+z)^2\frac{d^2}{dz^2}d_L-\frac{1}{2}(1+z)\frac{d}{dz}d_L+\frac{1}{2}d_L\right]~.
\eeq Assuming that the observed $d_L$ is described by the
corresponding theoretical relation for a $\lambda$CDM universe
\beq\label{LCDMdL}
d_L=\frac{1+z}{H^{\lambda\rm{CDM}}_0}\int\limits_1^{1+z}\frac{dx}{\sqrt{\alpha
x^3+\beta}}~, \eeq where $\alpha\simeq0.3$, $\beta \simeq 0.7$ and
$H^{\lambda\rm{CDM}}_0=70$ km/s/Mpc, we have \beq\label{source}
\frac{1}{H_0}\frac{d^2}{d\etas^2}\left[\frac{d_L}{1+z}\right] =
\frac{3}{2}\frac{H_0}{H^{\lambda\rm{CDM}}_0}\frac{(1+z)^2\beta}{\left(\alpha(1+z)^3+\beta\right)^{3/2}}~.
\eeq

\begin{figure}
\begin{minipage}[t]{5cm}\includegraphics[width=5cm]{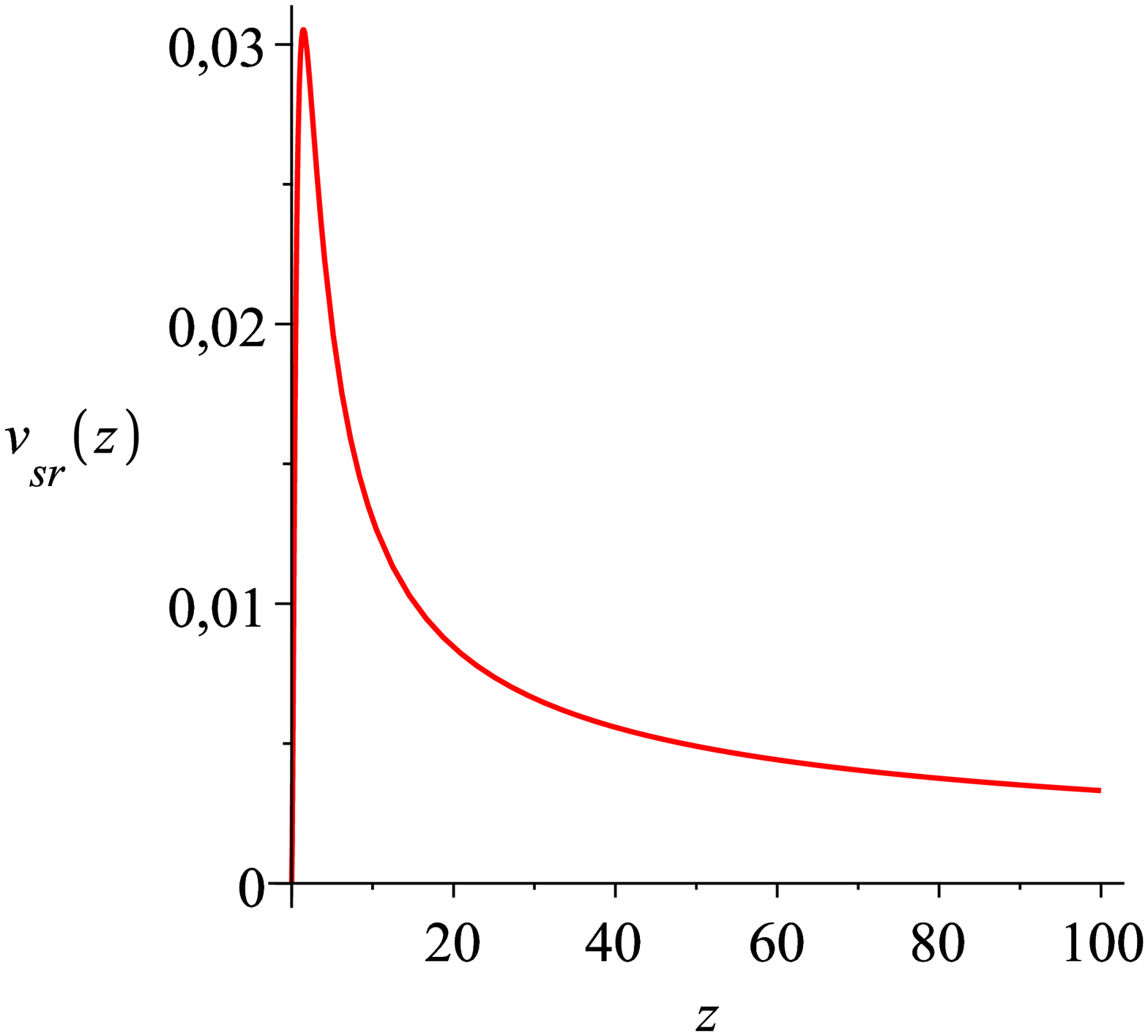}\end{minipage}\hspace{2cm}
\begin{minipage}[t]{5cm}\includegraphics[width=5cm]{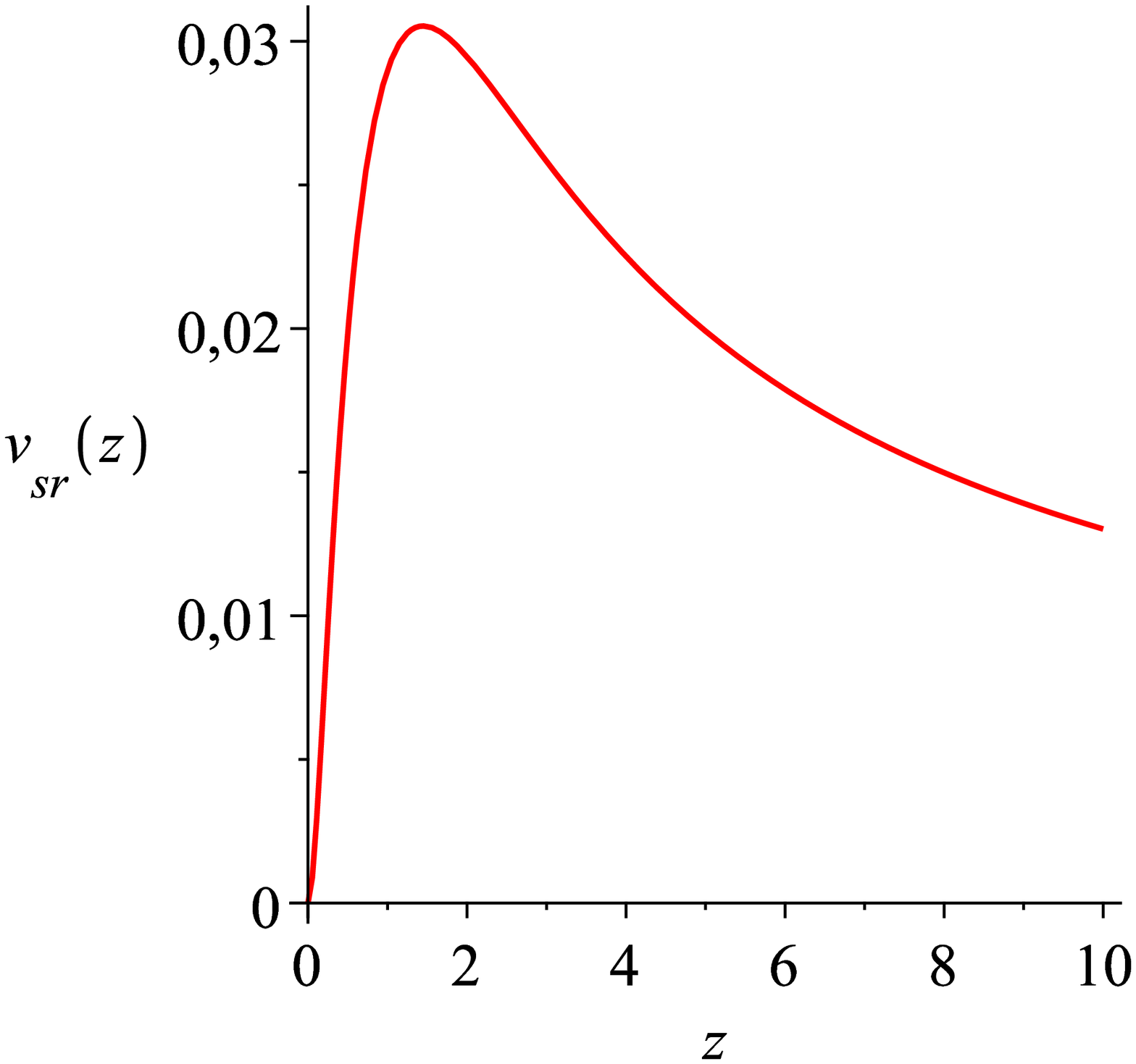}\end{minipage} \\
\vspace{0.5cm}
\begin{minipage}[t]{5cm}\includegraphics[width=5cm]{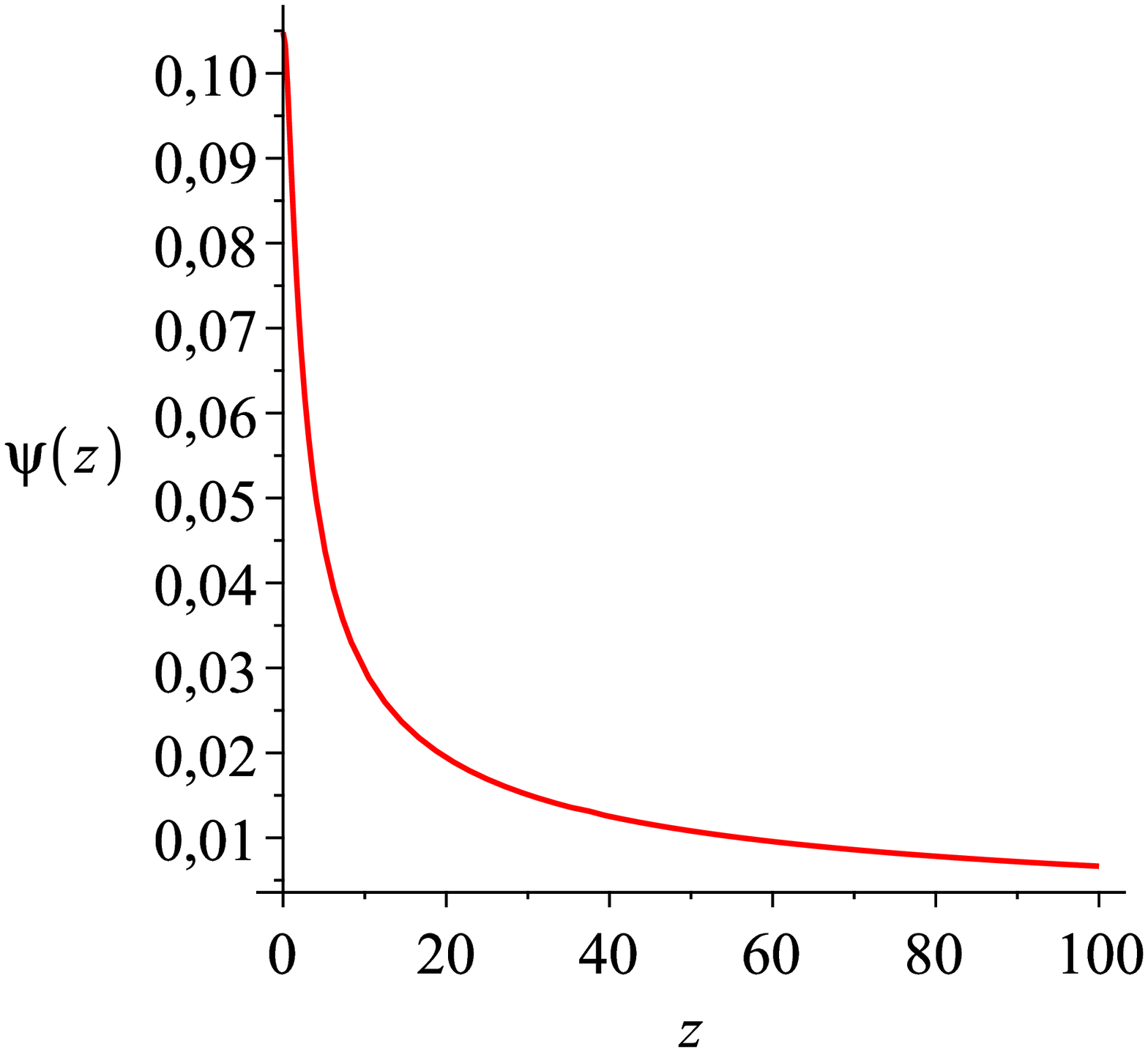}\end{minipage}\hspace{2cm}
\begin{minipage}[t]{5cm}\includegraphics[width=5cm]{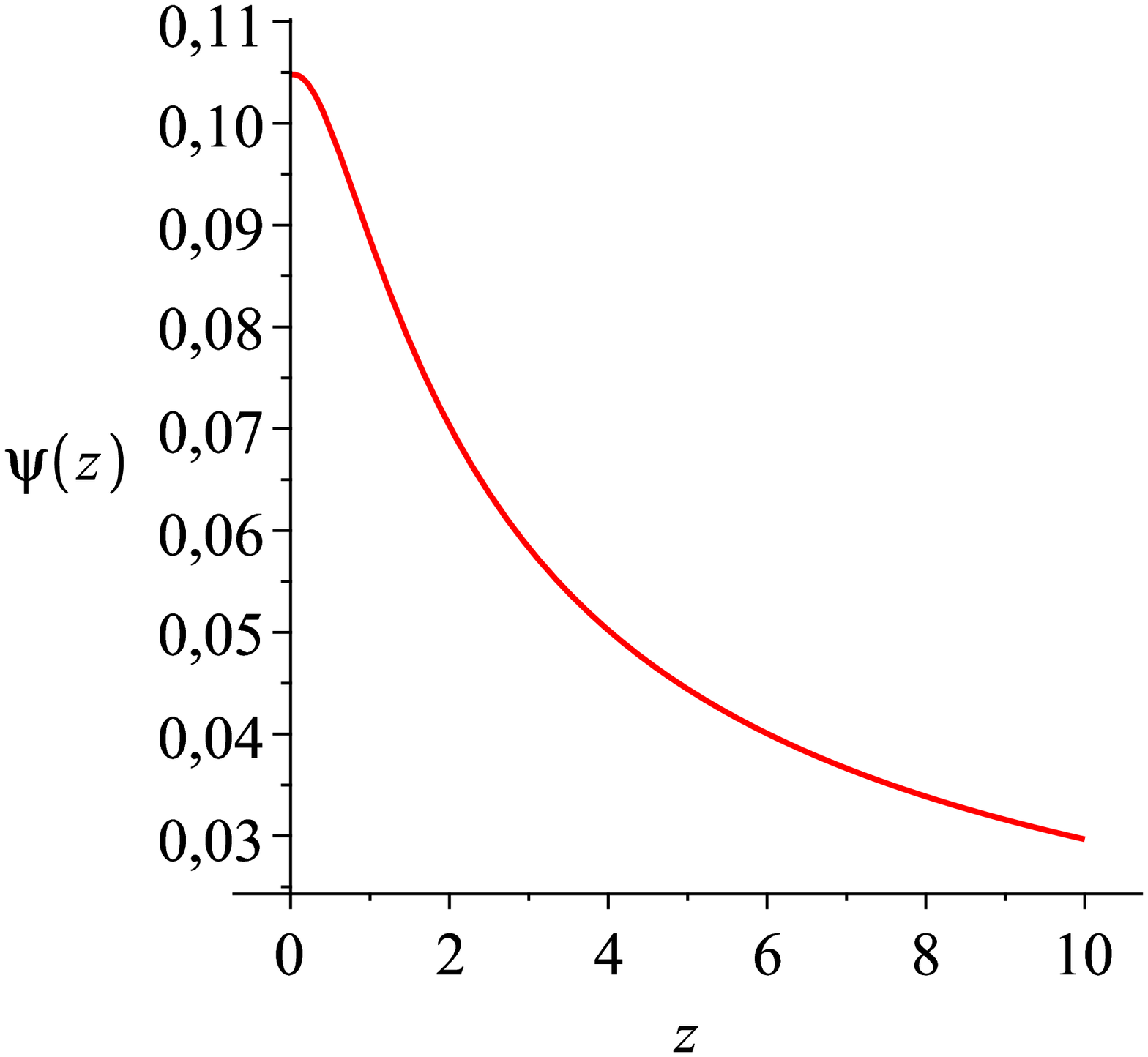}\end{minipage}
\caption{The velocity field $v_{sr}(z)$ and the gravitional
potential $\psi(z)$ of the perturbed FRW universe. The plots on the
right are close-ups of those on the left for redshifts $z<10$.}
\label{fig1}
\end{figure}

In Fig.\ \ref{fig1} we plot the solution to the equation of motion
for the velocity field, given by Eq.\ (\ref{basic2}), that gives
rise to the luminosity distance given in Eq.\ (\ref{source}). We
also construct the gravitational potential, assuming that $\psi$ is
zero at large redshifts. We see that a spherically symmetric linear
perturbation can indeed explain the observed acceleration with small
velocities and small gravitational potentials. Note that the value
$H_0$ in the hypothetical $\Omega =1$ dust universe is not really an
observable since the model is unlikely to fit all cosmological data.
For the purposes of the present approach, it is an adjustable
parameter. Here we have chosen $H_0=50$ km/s/Mpc for illustrative
purposes only. As is obvious from the gravitational potential of
Fig.\ \ref{fig1}, physically the perturbation corresponds to a
spherical underdensity around us.

\begin{figure}
\begin{minipage}[t]{5cm}\includegraphics[width=5cm]{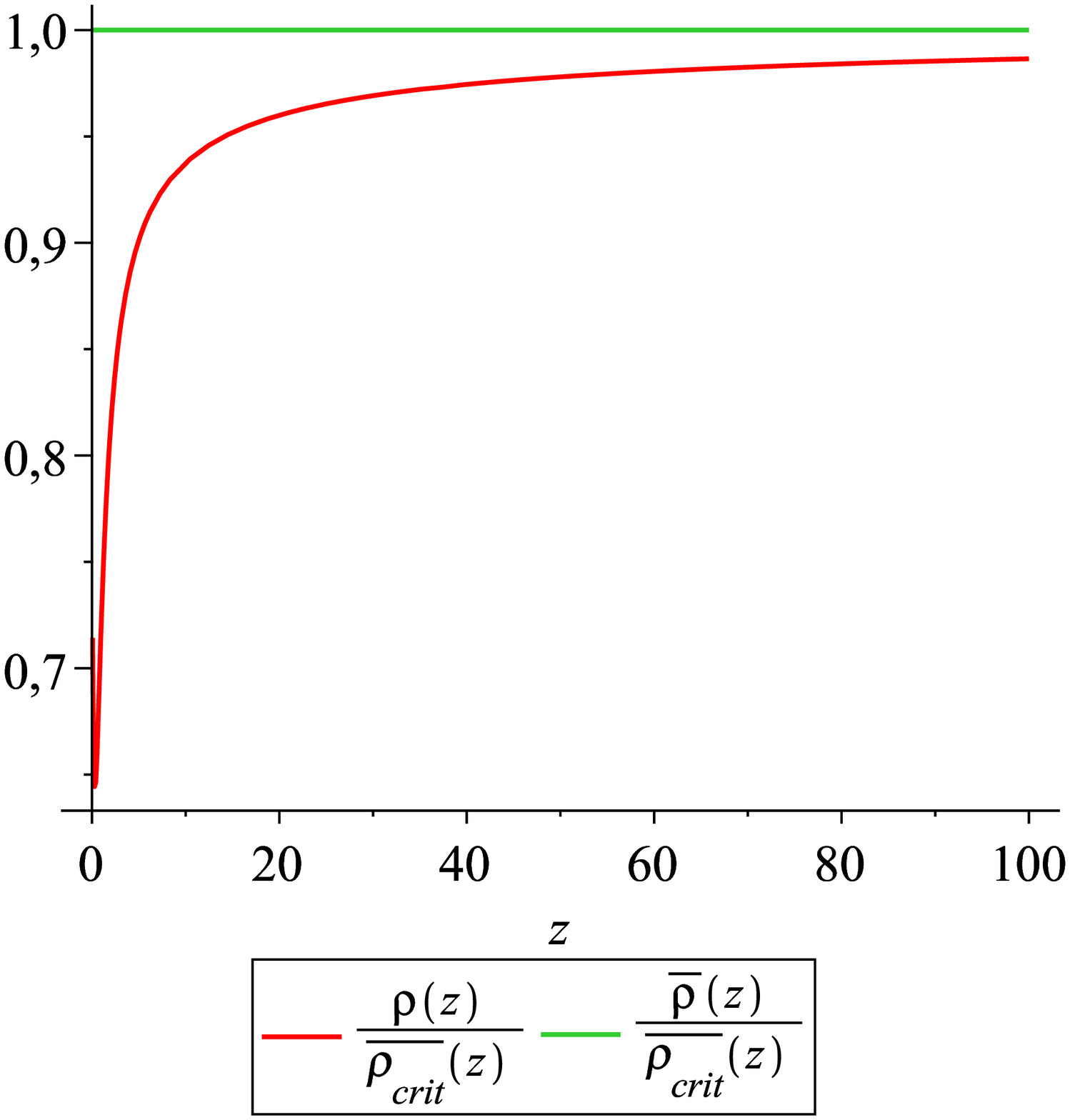}\end{minipage}\hspace{2cm}
\begin{minipage}[t]{5cm}\includegraphics[width=5cm]{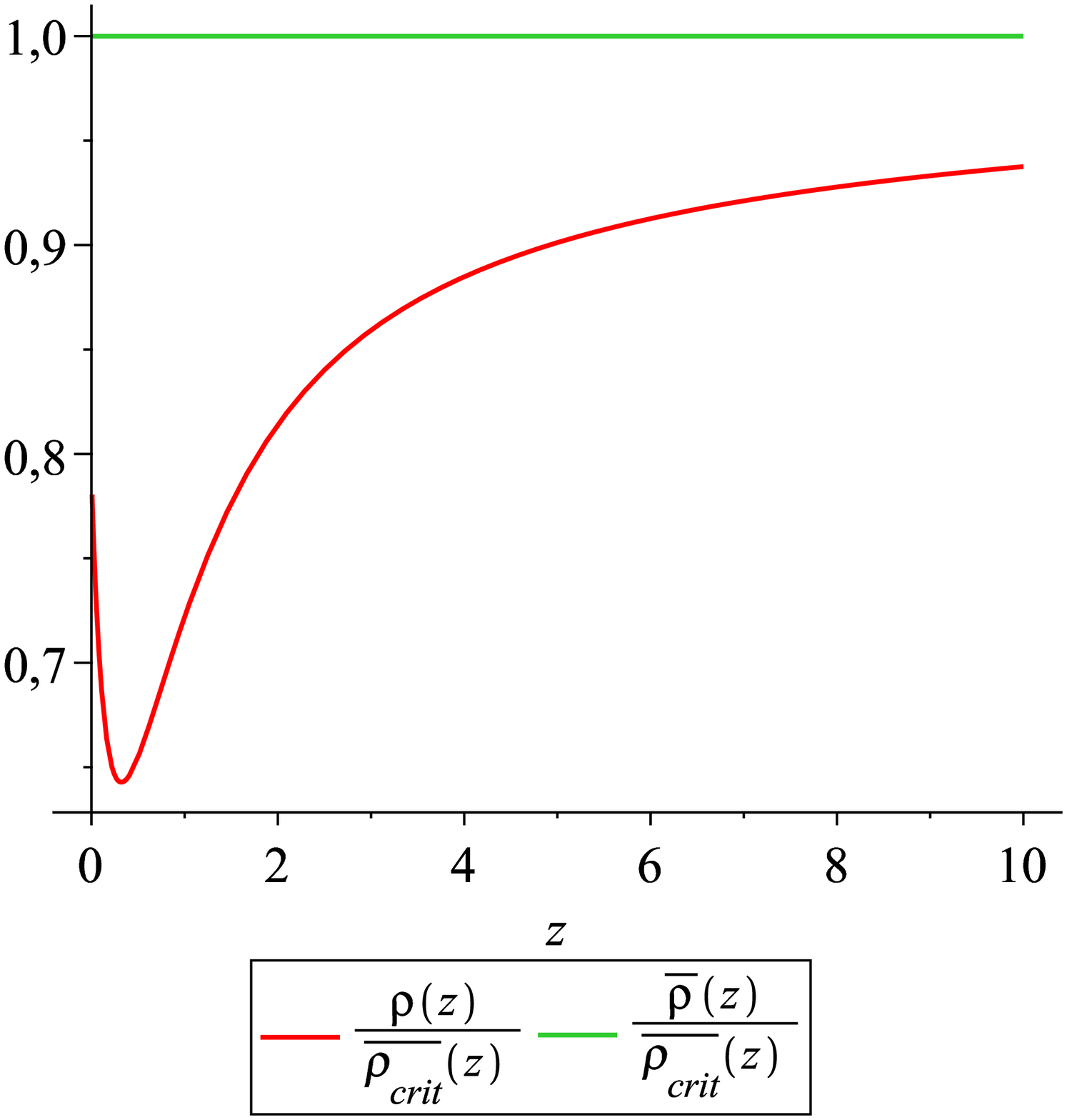}\end{minipage}
\caption{The density profile $\rho(z)/\bar{\rho}_{\rm{crit}}(z)$
(the red line). The green line corresponds to the background
Einstein-de Sitter $\bar{\rho}(z)/\bar{\rho}_{\rm{crit}}(z)=1$. The
plot on the right is a close-up of that on the left for redshifts
$z<10$.} \label{fig2}
\end{figure}

The density profile on the light cone can be immediately obtained
via the poisson equation \beq 4\pi a^{2} G\delta\rho =
\nabla^2\psi-3\mc{H}^2\psi~. \eeq Using Eqs.\ (\ref{nablasquared})
and (\ref{etazeta}) we thus have for the total matter density
$\rho=\bar{\rho}+\delta\rho$, where $\bar{\rho}=3H_0^2/(8 \pi
G)(1+z)^3$, along the light cone:
\begin{equation}\label{rho}
8\pi G\rho(z)=3H_0^2(1+z)^3\Big[1-\frac{1}{2}\frac{(1+{1}/
{\sqrt{1+z}})}{(1-{1}/{\sqrt{1+z}})}v_{sr}(z)-(1+z)
\frac{dv_{sr}(z)}{dz}-2\psi(z)\Big]~.
\end{equation}
In Fig.\ \ref{fig2} we plot $\rho(z)/\bar{\rho}_{\rm{crit}}(z)$,
where $\bar{\rho}_{\rm{crit}}(z)=\bar{\rho}(z)=3H_0^2/(8 \pi
G)(1+z)^3$.

\section{Transformation to LTB coordinates}
\label{LTBlink}

Let us now turn to LTB models. The LTB line-element with the spatial
origin at the symmetry center reads as
\begin{equation}\label{LTBmetric}
ds^2 = - d\tilde{t}^2 + \frac{(A'(\tilde{t},
\tilde{r}))^2}{1-k(\tilde{r})}d\tilde{r}^2 + A^{2}(\tilde{t},
\tilde{r}) \left( d\theta^2 + \sin^2 \theta d\varphi^2 \right)~,
\end{equation}
where $A(\tilde{t}, \tilde{r})$ is the scale function having both
temporal and spatial dependence and $k(\tilde{r})$ is a function
associated with the curvature of $\tilde{t}={\rm{const.}}$
hypersurfaces. Both $\tilde{t}$ and $A$ have a direct physical
meaning, the first one being the proper time of observers comoving
with the dust and the second the physical radius of a sphere with
area $4\pi A^2$. The coordinate $\tilde{r}$ is an arbitrary label,
devoid of physical meaning. We use the following shorthand notations
for the partial derivatives:
$'\equiv\frac{\partial}{\partial\tilde{r}}$ and
$\dot{}\equiv\frac{\partial}{\partial\tilde{t}}$. The LTB metric
(\ref{LTBmetric}) is an exact solution of the Einstein equations and
the perfectly homogeneous FRW model is a special case, obtained in
the limit: $A(\tilde{t}, \tilde{r}) \rightarrow
a(\tilde{t})\tilde{r}$ and $k(\tilde{r})\rightarrow k\tilde{r}^2$,
where $a(\tilde{t})$ is the FRW scale factor and $k$ is the
curvature constant.

It has been shown that under certain conditions, the LTB metric
(\ref{LTBmetric}) can be brought to the perturbed FRW form
(\ref{pertmetric}) via the following nonlinear gauge transformation
\cite{Paranjape:2008ai}, \cite{VanAcoleyen:2008cy}:
\begin{eqnarray}
 r&=& \frac{A(\tilde{t}, \tilde{r})}{a(\tilde{t})}(1+\xi(\tilde{t}, \tilde{r})) \label{coordtrans1} \\
 t &=& \tilde{t}+\xi^0(\tilde{t}, \tilde{r})~, \label{coordtrans2}
\end{eqnarray}
where $a(\tilde{t}) \equiv (\tilde{t}/\tilde{t}_0)^{2/3}$ is a
fictitious FRW scale factor corresponding to an $\Omega=1$ matter
dominated universe. If the LTB model is to be close the the FRW, the
functions $\xi(\tilde{t}, \tilde{r})$ and $\xi^0(\tilde{t},
\tilde{r})$ must satisfy $|\xi|, |\xi^0H| \ll 1$. We will
demonstrate that these conditions are indeed satisfied for our case
on our past light cone.

Before proceeding we should note that Eq.\ (\ref{coordtrans2})
provides a relation between the clock times of the LTB and
(fictitious) FRW observers. Their difference can be intuitively
understood in two ways: In a local description, the LTB coordinates
are synchronous, meaning that they are defined by the world-lines of
the dust particles which, since pressure is absent, are geodesics.
Therefore, their clocks will tick faster than those of any other
non-geodesic coordinate system such as the FRW coordinates. The same
conclusion is reached in a global description where the LTB
spacetime describes a physical underdensity around our position,
meaning that the LTB clocks will tick faster with respect to the FRW
coordinates in this case. This effect will be more pronounced as we
move to the centre and thus $H\xi^0$ is expected to decrease with
decreasing redshift. Eq.\ (\ref{coordtrans1}) defines the FRW
distance $r$ in terms of the physical distance $A$, a prescribed FRW
scale factor $a(\tilde{t})\equiv
\left(\tilde{t}/\tilde{t}_0\right)^{2/3}$ and a function $\xi$ which
is chosen to ensure that on the light cone $r$ takes the standard
FRW form in terms of redshift: \beq\label{xi-choice}
\frac{A(\tilde{t}(z), \tilde{r}(z))}{a(z)}(1+\xi(\tilde{t}(z),
\tilde{r}(z))) =2H_0^{-1}(1-1/{\sqrt{1+z}})~. \eeq Furthermore, we
choose the arbitrary $\tilde{r}$ such that, on the light cone:
\beq\label{tilder-choice}
\partial_{\tilde{r}}A(z)=a(z)~.
\eeq

Let us now determine $H\xi^0$ and $\xi$. In the LTB model, the
angular diameter distance is related to the metric simply by
\cite{Krasinski2} \beq\label{angdist}d_A(z)=A(\tilde{r}(z),
\tilde{t}(z))=\frac{d_L(z)}{(1+z)^2}~, \eeq where the last equality
is due to Etherington's theorem \cite{Etherington, Ellis} stating
that for geodesic light in any spacetime the luminosity and angular
diameter distances are related through $d_L=(1+z)^2d_A$. If the
perturbed FRW and LTB metrics are to describe the same spacetime,
just written in different coordinates, the observable
distance-redshift relations should be the same:
$d_A^{\rm{LTB}}=d_A^{\lambda\rm{CDM}}$. So, by having the observable
distance-redshift relations we can readily read off
$A(\tilde{t},\tilde{r})$ along the past light cone of the LTB model,
that is $A(z)$. $\xi$ can then be determined directly from
(\ref{xi-choice}).

Comparing the angular parts of metrics (\ref{pertmetric}) and
(\ref{LTBmetric}) we see that, to leading order: \beq\label{A(r,t)}
A(\tilde{t}, \tilde{r}) \simeq a(t)r\left[1-\psi(t,r)\right] \simeq
a(\tilde{t})r\left[1-\psi(\tilde{t},
\tilde{r})+H(\tilde{t})\xi^0(\tilde{t}, \tilde{r})\right]~, \eeq
where we have used (\ref{coordtrans2}). In the previous Section we
determined the $v(z)$ (or $\psi(z)$) profile giving the $\Lambda$CDM
luminosity distance in the perturbed FRW dust universe. Using Eqs.\
(\ref{coordtrans1}), (\ref{xi-choice}) and (\ref{angdist}), we can
thus obtain the small quantity $H\xi^0$ from Eq.\ (\ref{A(r,t)})
along our line of sight:
\begin{eqnarray}\label{ksinot}
&d_A(z)& = \frac{2}{H_0(1+z)}\Big(1-\frac{1}{\sqrt{1+z}}\Big)\Big(1-\psi(z)+H(z)\xi^0(z)\Big) \\
&\Longrightarrow& H(z)\xi^0(z) =
\frac{d_A(z)(1+z)H_0}{2(1-\frac{1}{\sqrt{1+z}})}+\psi(z)-1~,
\end{eqnarray}
where $d_A$ is obtained from Eq.\ (\ref{LCDMdL})
($d_A=(1+z)^{-2}d_L$).

\begin{figure}
   \begin{minipage}[t]{5cm}\includegraphics[width=5cm]{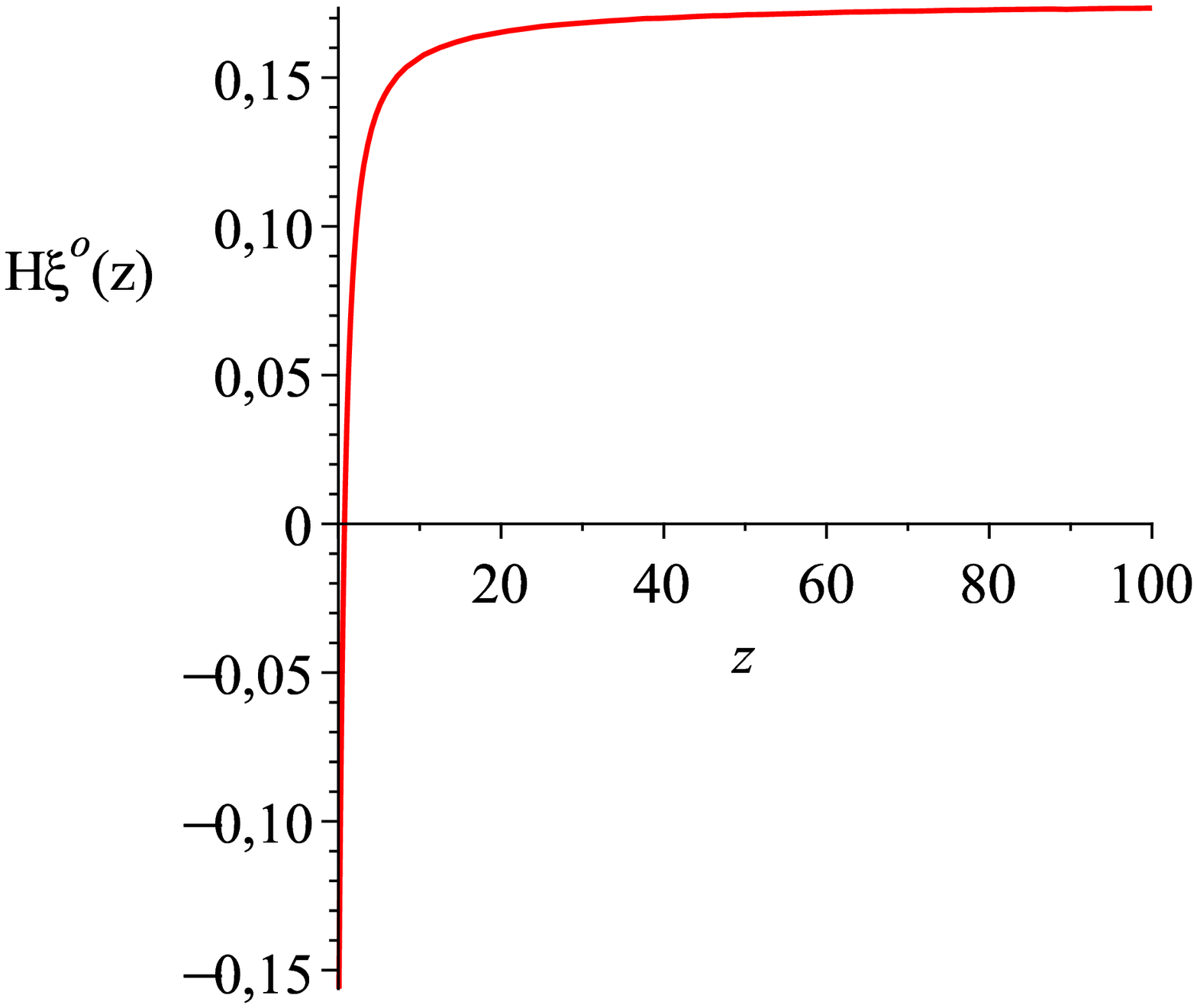}\end{minipage}\hspace{2cm}
   \begin{minipage}[t]{5cm}\includegraphics[width=5cm]{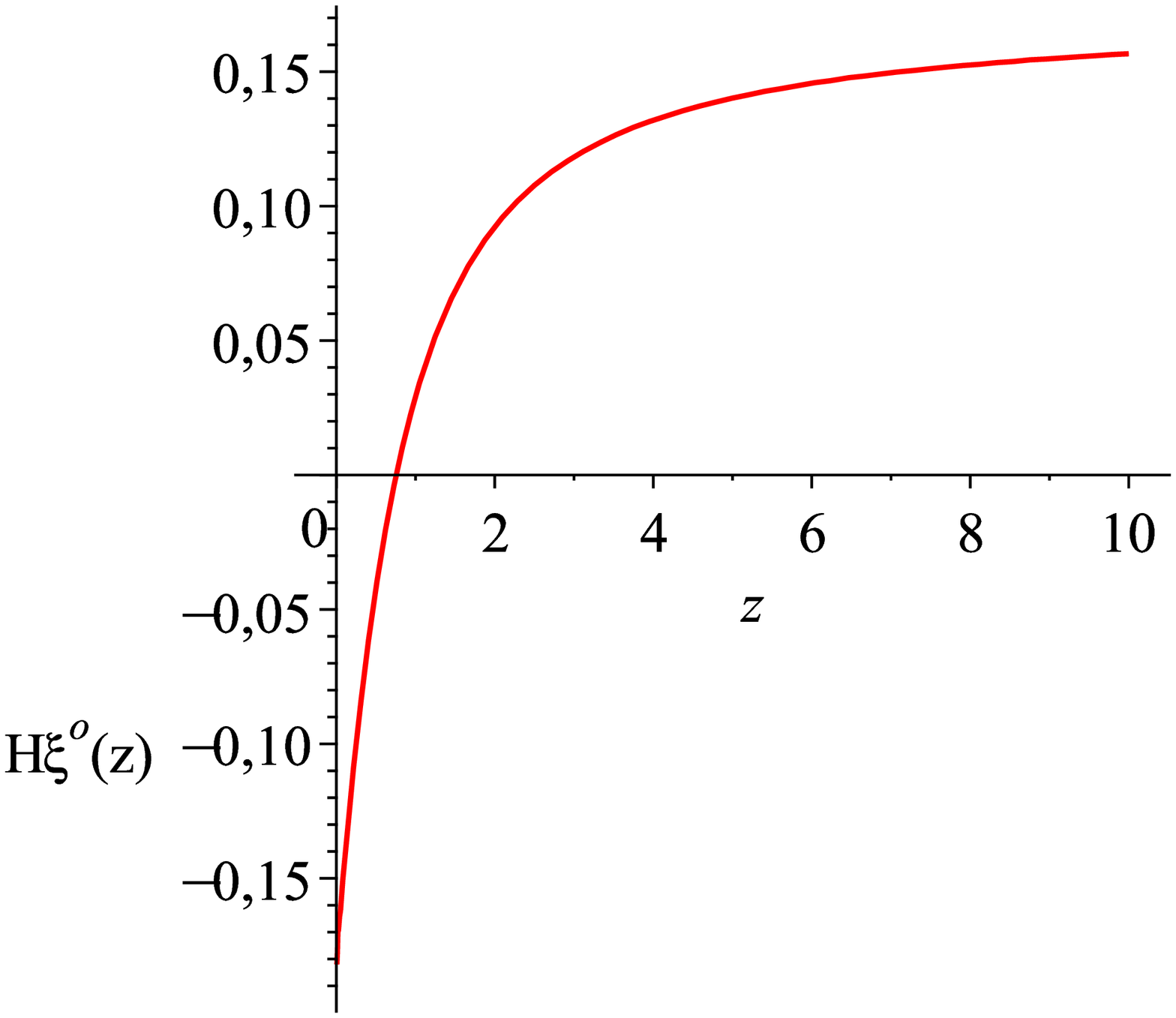}\end{minipage} \\
   \vspace{0.5cm}
   \begin{minipage}[t]{5cm}\includegraphics[width=5cm]{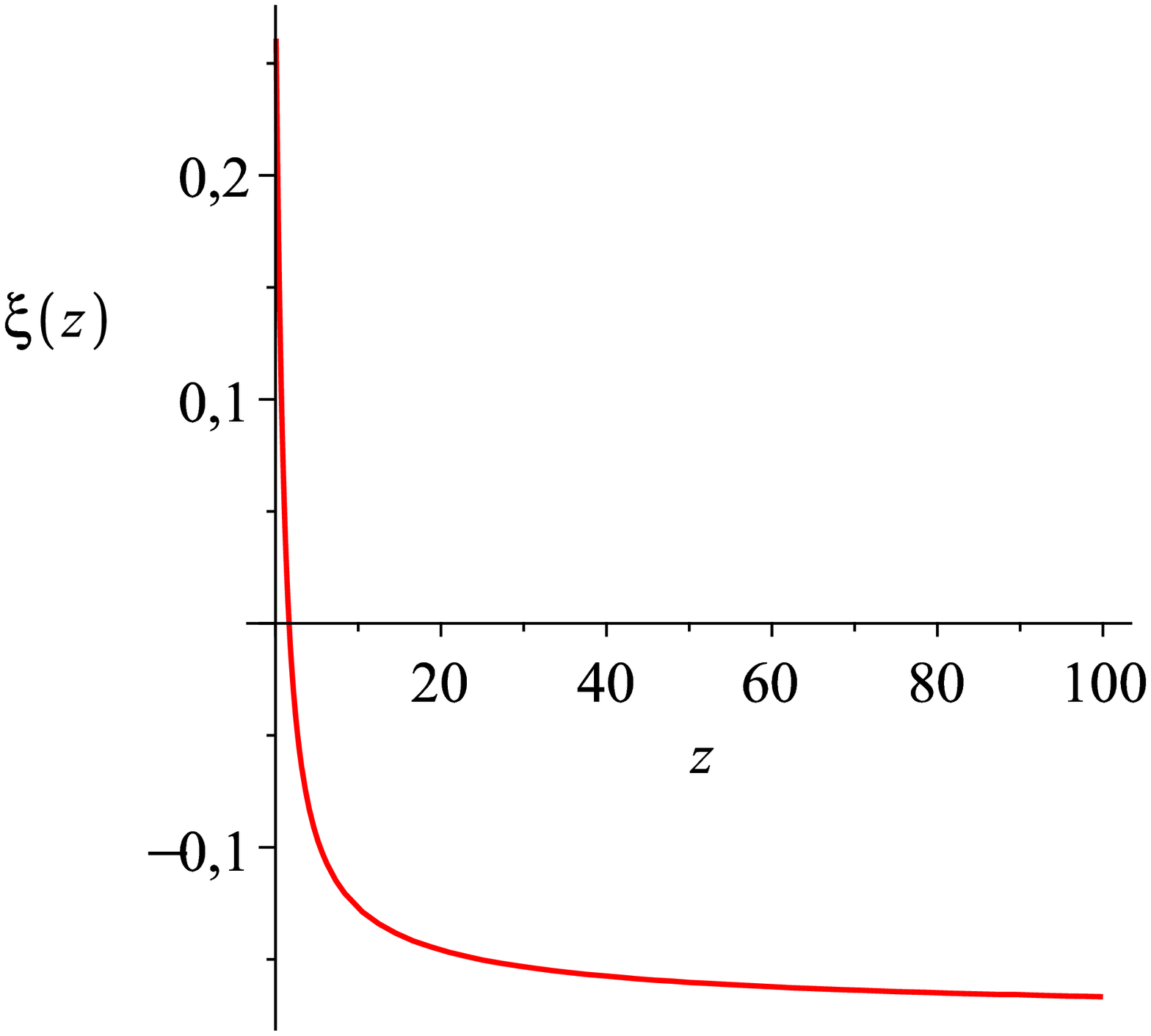}\end{minipage}\hspace{2cm}
   \begin{minipage}[t]{5cm}\includegraphics[width=5cm]{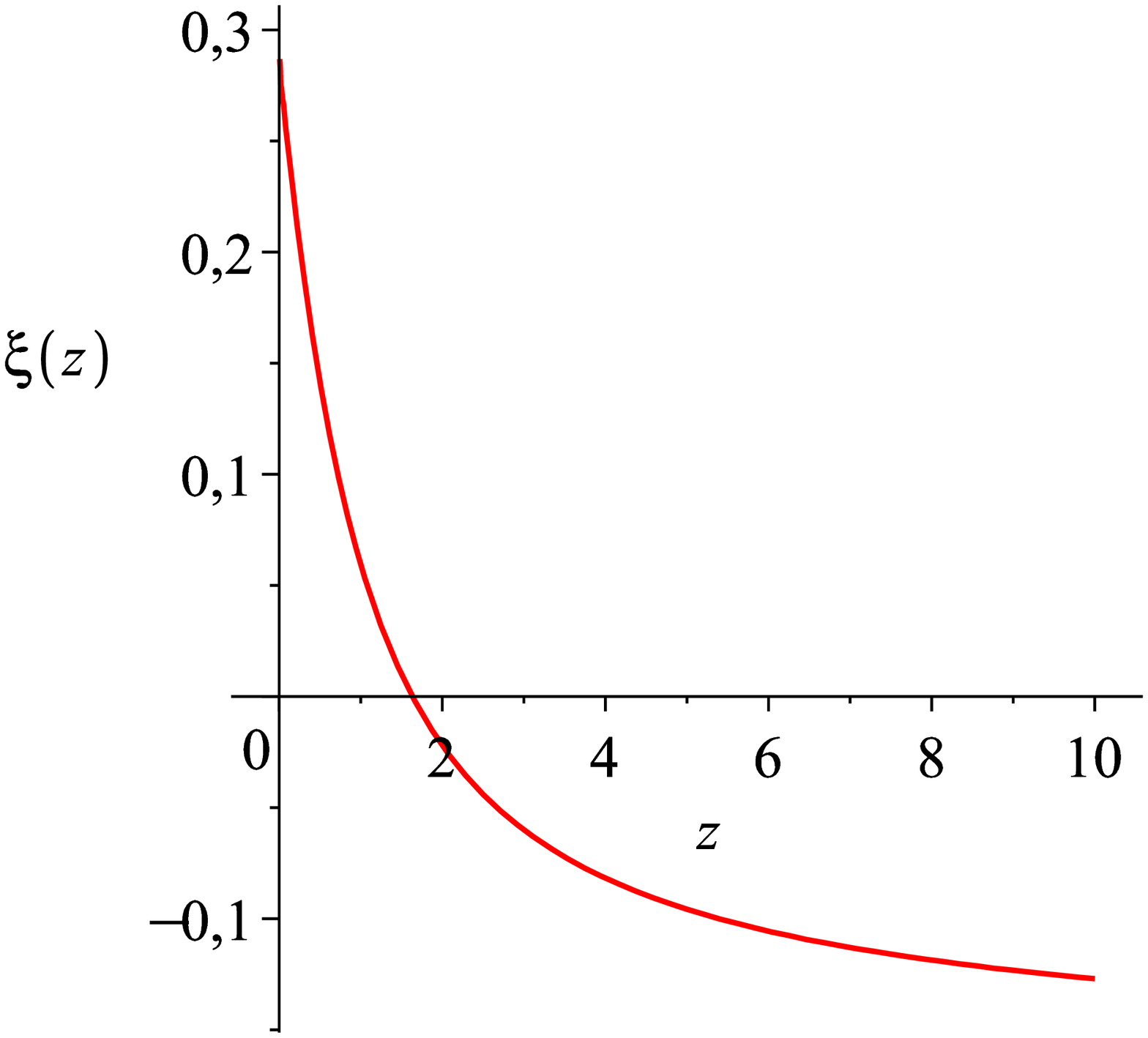}\end{minipage}
   \caption{The functions $H\xi^0(z)$ and $\xi(z)$ relating the perturbed FRW and LTB clock times. The plots on the
right are close-ups of those on the left for redshifts
$z<10$.}\label{fig3}
\end{figure}

The outcome is shown in Fig.\ \ref{fig3}, where we display the shift
between $t$ and $\tilde t$ in units of the inverse Hubble rate as
well as $\xi$. As can be seen both $H\xi^0$ and $\xi$ are less than
one. In principle, we could have kept higher orders in our
expansions, but other than changing the numerical values of
$\xi^0(z)$ by {\cal O(30)\%}, we do not expect any qualitative
change in this picture. Thus, the conclusion is that an LTB model
fitting the SNIa data can be mapped into a perturbed FRW model by a
nonlinear coordinate transformation of the form (\ref{coordtrans1}),
(\ref{coordtrans2}), \emph{ at least on the light cone}.

In the literature, there has been discussion about the so called
"inverse-problem", that is, finding a map from a given luminosity
distance to the corresponding LTB model
\cite{Iguchi:2001sq,Chung:2006xh,Yoo:2008su}. In fact, in
\cite{Yoo:2008su} an LTB model, whose $d_L - z$ relation is
equivalent to that of the $\Lambda$CDM model was constructed. Not so
surprisingly, the resulting LTB model was found to be such that
there is a void whose symmetry center is at the observer's position
- assuming the early universe was homogeneous. As expected, our
results are consistent with this: In the previous Section we found
that the perturbed FRW fitting the $\Lambda$CDM $d_L - z$ curve
contains a void around our location, given the universe is
asymptotically Einstein-de Sitter. Since the LTB and perturbed FRW
are the same spacetime, usually just written in different
coordinates, the LTB model must also represent this void. So
although we have not explicitly calculated the quantitative
properties of the resulting LTB model, the physical content is known
from a qualitative point of view and consistent with
\cite{Yoo:2008su}, as it should.

Before summarizing our results, a few comments on the paper
\cite{Kolb:2008bn} are in place, as their results seem to contradict
ours. As a toy model, they consider an LTB model consistent with the
SNIa data with a large (1 Gpc) void in a universe approaching
spatially to Einstein-de Sitter. They find that such models cannot
be described in terms of a conformal Newtonian metric perturbed
about an FRW background during the entire evolution without large
peculiar velocities. However, our result is that the SNIa data can
indeed be explained by the perturbed FRW model as well as by an LTB
model, with these two descriptions being equivalent and with the
peculiar velocity remaining small. Correspondingly, we find that the
shifts between FRW and LTB coordinates are also small, in agreement
with the analysis of \cite{VanAcoleyen:2008cy} (see also
\cite{Paranjape:2008jc}). In their analysis these authors include
nonlinear but leading-order terms in the velocities which are absent
from \cite{Kolb:2008bn}. Indeed, since the solution is in the
nonlinear regime, one would have to employ the full nonlinear
transformation. However, more important is the fact that the
peculiar velocities of the model considered in \cite{Kolb:2008bn}
become large at redshifts $z\gtrsim 500$, when the void represents a
nonlinear superhorizon density fluctuation, so the model is not even
in principle describable within the weak field description of
Newtonian gauge. This means that there actually is no discrepancy
between their consideration and ours: one can achieve large
deviations from the perturbed FRW picture, but only by choosing
initial conditions which seem observationally unrealistic. Indeed,
when confronted with the CMB observations, a universe dominated by a
Gpc sized void seems unlikely.

\section{Discussion}

We have established that the FRW $\Omega = 1$ dust universe with
spherically symmetric perturbations in the gravitational potential
$\psi(z)$ can fit the SNIa data. We found that physically the
perturbation corresponds to a void around us. We do not pretend to
know what could be the origin of the perturbation or the associated
velocity field, and in fact it is rather obvious that the model as
such is likely to be way too simplistic to explain all the relevant
cosmological data. However, the interesting point here is that the
SNIa data can be explained by small deviations from FRW; both
$\psi(z)$ and the velocity field ${\bf v}(z)$ remain perturbative at
all distance scales. Moreover, such a perturbative approach seems to
capture all the essential features of the data. As we demonstrated,
if the non-perturbative, spherically symmetric LTB model fits the
SNIa data, it can be mapped into the perturbed FRW model by a
nonlinear coordinate transformation, at least on the light cone.

Our result seem to contradict \cite{Kolb:2008bn}, where the claim is
that the LTB metric cannot be expressed as a perturbed conformal
Newtonian metric during the entire evolution of the universe without
the existence of a peculiar velocity larger than the speed of light.
However, they use a 1 Gpc LTB void model as a tool to demonstrate
this and indeed, such case indicates the breakdown of the
perturbative description: their model approaches Einstein-de Sitter
only spatially and there were strong density contrasts in the early
universe. The question is then whether such large nonlinear
inhomogeneities are realistic. Observations seem to suggest that the
inhomogeneities at the last scattering surface do not lead to such
large underdense regions. Our model approaches asymptotically to
Einstein-de Sitter along the past light cone, which implies that the
early universe was very homogeneous and as our calculations
explicitly demonstrate, the explanation of the supernovae data does
not require a breakdown of the weak field description. As shown in
Fig.\ \ref{fig1}, the peculiar velocity remains small during the
entire evolution along the past light cone and the mapping from LTB
to perturbed FRW is possible. This is consistent with the results of
\cite{Paranjape:2008jc,VanAcoleyen:2008cy}.

We should like to stress that in fact we have not specified the LTB
model quantitatively. That is, we have neither determined the LTB
curvature function $k(\tilde{r})$ nor the mass function
$F(\tilde{r})$, nor the entire evolution of $A(\tilde{t},
\tilde{r})$ inside our past light cone. However, the underdense
bubble around us, represented by the perturbed FRW model, has to be
also the physical content of the resulting LTB model from a
qualitative point of view since, as we verified, the LTB and
perturbed FRW are the same spacetime on our past light cone. In
principle, we should have all the information to unambiguously
determine also the quantitative properties of that LTB model: Our
physical situation is such that the universe approaches
asymptotically Einstein-de Sitter along the past light cone, so from
this consideration it is obvious that the LTB model contains only
growing mode inhomogeneities, so that the early universe was
homogeneous and the Big Bang was homogeneous or the age of the
universe is independent of the spatial coordinates:
$\tilde{t}_0(\tilde{r})=\rm{constant}$. This requirement sets a
constraint between the two boundary condition functions of the LTB
model, thus leaving only one free function and a free parameter
$\tilde{t}_0$, just like in the perturbed FRW case - the potential
$\psi(z)$ (or alternatively, the velocity field ${\bf v}(z)$) and a
free parameter $H_0$. However, our interest here was not to
concentrate on the exact model building. In fact, the issue of
finding an LTB model that fits to a given $d_L - z$ curve has been
previously addressed by a few authors
\cite{Mustapha:1998jb,Iguchi:2001sq,Chung:2006xh,Yoo:2008su} and
indeed, the physical content of the LTB model described here is
consistent with the result obtained in \cite{Yoo:2008su}, where an
LTB model was constructed by demanding its luminosity distance to be
equivalent with the one in the $\Lambda$CDM model - their result is
a void, the center of which we occupy. We hope to return to the
issue of quantitatively determining the LTB model from luminosity
distance data in future work. It will also be interesting to see
whether the perturbative nature of the model persists inside the
past light cone.

We thus conclude that the SNIa data does not require a deviation
from the Copernican Principle that is non-perturbative, at least if
we limit ourselves to spherically symmetric perturbations. Indeed,
it would be of some interest to generalize the present approach to
non-spherical perturbations; whether this can be done in practice
remains to be seen.

\vskip31pt

\centerline{\textbf{Acknowledgements}} \vskip10pt This work was
partly supported by the EU 6th Framework Marie Curie Research and
Training network "UniverseNet" (MRTN-CT-2006-035863) and partly by
Academy of Finland grant 114419. MM is supported by the Graduate
School in Particle and Nuclear Physics. MM wishes to thank Teppo
Mattsson for useful discussions.

\end{document}